\documentclass[a4paper]{jpconf}
\usepackage{graphicx}

\def\beq{\begin{equation}}
\def\eeq{\end{equation}}
\def\bea{\begin{eqnarray}}
\def\eea{\end{eqnarray}}
\def \lsim{\mathrel{\vcenter
     {\hbox{$<$}\nointerlineskip\hbox{$\sim$}}}}
\def \gsim{\mathrel{\vcenter
     {\hbox{$>$}\nointerlineskip\hbox{$\sim$}}}}

\def\a{\alpha}
\def\b{\beta}

\def\r{\rho}

\begin{document}
\title{Non standard neutrino interactions at LEP2 and the LHC}

\author{Sacha Davidson$^1$ and Veronica Sanz$^2$}

\address{$^1$ IPNL, Universit\'e de Lyon, Universit\'e Lyon 1, CNRS/IN2P3, 
4 rue E. Fermi 69622 Villeurbanne cedex, France  }

\address{$^2$ Department of Physics and Astronomy, York University, Toronto, ON, Canada }

\ead{$^1$ s.davidson@ipnl.in2p3.fr}
\ead{$^2$  vsanz@yorku.ca }

\begin{abstract}
We consider  Non-Standard neutrino Interactions (NSI)
connecting  two neutrinos with two  first-generation fermions ($e, u$ or $d$),
which we assume to arise at
 dimension eight due to New Physics.
The  coefficient is normalised as $4 \varepsilon G_F/\sqrt{2}$.
We explore  signatures  of NSI-on-electrons at
LEP2, and of NSI-on-quarks at the LHC,
treating the NSI as contact interactions
at both energies.
In models where the coefficients of dangerous  dimension six operators
are suppressed by  cancellations,  
LEP2  provides interesting 
bounds on NSI operators 
($\varepsilon \lsim 10^{-2} - 10^{-3}$),
which arise because  $\sqrt{s} \sim 200$ GeV, 
and the cancellation  applied at zero
momentum transfer.
At the LHC, we use  the Equivalence Theorem,
which relates the longitudinal $W$ to
the Higgs, to estimate the  rate for  
  $\overline{q} q  \to W^+W^- e_\alpha^+ e_\beta^-$
induced by NSI. We find that the cross-section
is small, but that the outgoing particles
have very high $p_T > 400$ GeV, which
reduces the issue of backgrounds. 
In a conservative scenario, we find that the  LHC at 14 TeV and with 100 fb$^{-1}$ of data 
would have a
  sensitivity   to $\varepsilon \gsim 3 \times  10^{-3}$. \\$~$\\
Contribution  to
NUFACT 11, XIIIth International Workshop on Neutrino Factories, Super beams and Beta beams, 1-6 August 2011, CERN and University of Geneva
(Submitted to IOP conference series)
\end{abstract}

\section{Introduction}

Many extensions of
the Standard Model, such as Supersymmetry or
leptoquarks, naturally induce  low energy  contact
 interactions of the form:
\beq
\varepsilon_{\a \b}^{fX}  \frac{4 G_F}{\sqrt{2}} 
(\overline{f} \gamma^\r P_X f) (\overline{\nu}_\a \gamma_\r \nu_\b) 
~~~,
\label{eqndisc}
\eeq
where $f \in \{ u, d,e \}$ is a first generation 
charged Standard Model fermion, and
$\a,\b \in \{e, \mu, \tau\}$. These  are referred
to as Non Standard neutrino
Interactions (NSI) \cite{nuFact,slepton}, and can be
generated by $SU(3) \times SU(2) \times U(1)$ 
gauge invariant operators at dimension
 six or higher.  Future neutrino
facilities, such as a Neutrino Factory \cite{nuFact},
could be sensitive to such
interactions with 
 $\varepsilon \gsim 10^{-4}$.
Formalism and   current bounds \cite{Biggio:2009nt} on NSI are reviewed
 in the paper on which this proceedings is based \cite{SD}.

This  proceedings reports on  a preliminary  exploration \cite{SD}
of the complementarity 
of  current
collider experiments and future neutrino 
facilities  to
these neutral current operators. We focus 
on neutral current NSI induced at dimension eight,
from operators such as  
\beq
\label{eqn2}
\frac{1}{\Lambda_8^4}
(\overline{q} \gamma^\r P_L q) (\overline{H\ell }_\a \gamma_\r H \ell_\b) 
\eeq
where $q$ and $ \ell$ are SM doublets, and  $H$ is the Higgs.
Comparing with eqn  (\ref{eqndisc}) gives
$v^2/\Lambda_8^4  = \varepsilon/v^2  $, 
where $v = \langle H \rangle = 174$ GeV.
So $ \Lambda_8  \lsim 2 $ TeV, suggesting
that   new mediating particles,
if weakly coupled or  
in loops, are kinematically accessible
to the LHC. In this case, their
discovery prospects are 
model-dependent, and have been
widely studied\cite{tdrs}.  Here,
to retain some degree of model independence,
 we  consider effective operators at colliders
(LEP2, LHC).  
So  
we are assuming  heavy  New Physics with
$\gsim 1$ couplings.  We also assume 
that the New Physics does not generate
 ``dangerous''
 dimension six operators involving two
charged leptons instead of two neutrinos, because these are
strictly constrained \cite{slepton,Gavela}.

In section \ref{LEP2}, we argue that in
some models, such as those considered
by Gavela {\it et.al.} \cite{Gavela},
the dimension eight NSI interaction
of the form (\ref{eqndisc}) is accompanied by
dimension eight contact interactions
involving  charged leptons rather than neutrinos,
 with coefficients
$\sim s/\Lambda_8^4, (t-u)/\Lambda_8^4$, where $s$, $t$
and $u$
are the Mandelstam variables.

In section \ref{LHC}, we use the Equivalence Theorem
to replace $\langle H \rangle \nu_\a \to W^+ e_\a^-$, and study
the prospects for detecting $q \bar{q} \to W^+ W^- \ell_\a^+
\ell_\b^-$ at the LHC. Rough estimates suggest that
couplings $\gsim 1$ are required for 
 NSI  to function as dimension eight contact
interactions at LHC energies (assuming $\varepsilon \gsim 10^{-4}$).
Such contact interactions would induce few events ($\sigma (p p \to
W^=W^- \ell^+\ell^-)  \sim 10^{-3}$  fb $ \times (10^{-4}/\varepsilon)^2$),
 but at very high $p_T$ where Standard Model
backgrounds are negligeable. We estimate that
the LHC could be sensitive to 
$\varepsilon \gsim  3 \times 10^{-3}$.

\section{LEP2 Bounds on Dimension 8 Derivative Operators}
\label{LEP2}

Some NSI models, which  suppress dangerous
dimension six operators  via
 a cancellation  \cite{Gavela},
could give rise to 
four fermion contact interactions
 with coefficients $\propto \{ s,t,u \}/\Lambda_8^4$.
 Such contact interactions 
are subject to
LEP2 bounds. 
We present   estimates of the translation
of the LEP bounds to
the $\varepsilon$ coefficient of 
   NSI.

As an example  of a cancellation in
the coefficient  
of   a dangerous  dimension six 
operator, 
consider a model containing 
an  SU(2) doublet vector and scalar with large  masses $m_V$,$m_S$,
and couplings
\bea
h  (\overline{e} \ell) \tilde{S}_2^\dagger 
+ 
g (\overline{e^c} \gamma_\mu \ell){V}^\mu_2 
\eea
The 
exchange of these two bosons gives
the operators
\bea
- \frac{h^2}{m_S^2-t}  (\overline{e} \ell) (\overline{\ell} e)
+
 \frac{g^2}{m_V^2-u}(\overline{e^c} \gamma^\mu \ell) 
(\overline{\ell} \gamma_\mu e^c) 
=\left(
 -\frac{g^2}{m_V^2-u}
 +
 \frac{h^2}{2(m_S^2 -t)}
\right)
(\overline{\ell} \gamma^\mu \ell) (\overline{e} \gamma_\mu e) 
\eea
where $t$, $u$ are Mandelstam variables.
At zero momentum transfer, the
coefficient
can be cancelled 
by choosing  $g^2/m_V^2 = h^2/(2m_S^2)$.
However, at dimension 8, 
operators  arise such as 
\bea
  \frac{g^2(t/m_S^2-u/m_V^2)}{m_V^2}
(\overline{\ell} \gamma^\mu \ell) (\overline{e} \gamma_\mu e) 
\label{dim8}
\eea
We consider  two independent  
dimension eight operators, with coefficients
 $\propto s/\Lambda_8^4$ and $\propto (t-u)/\Lambda_8^4$,
and
that have  four visible fermion legs.

The  LEP2 experiments searched for dimension six 
 four fermion  contact interactions   
in the channels
$e^+e^- \to e^+ e^-,  \mu^+ \mu^-,  \tau^+ \tau^-,  \bar{q} q$
(where $q \in \{u,d,s,c,b \}$ is a light quark.
  The published  bounds
\cite{bourilkov,ALEPH,OPAL}  can be
translated  (to within factors of 2, see \cite{SD}) to
 bounds on  dimension eight operators, $\propto s/\Lambda_8^4$ or
$(t-u)/\Lambda_8^4$, with the same external legs. 
 If the NP inducing
the derivative operators has
${\cal O}(1)$ couplings to the Higgs, 
then  bounds
  on the $\varepsilon$ coefficient of  the 
dimension eight NSI  operator, can be obtained
by taking $ s \sim v^2$. See \cite{SD} for details.

\begin{figure}[h]
\begin{minipage}{16pc}
\includegraphics[width=16pc]{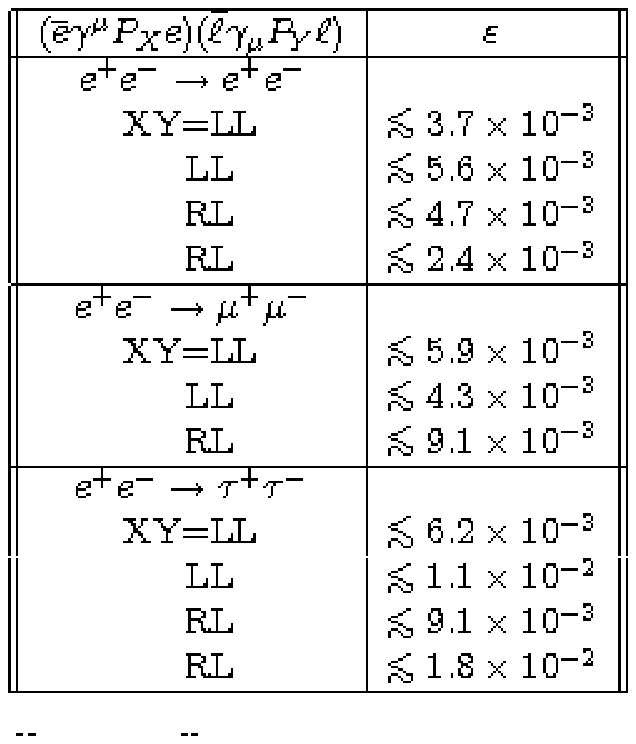}
\end{minipage}\hspace{2pc}%
\begin{minipage}{16pc}
\includegraphics[width=16pc]{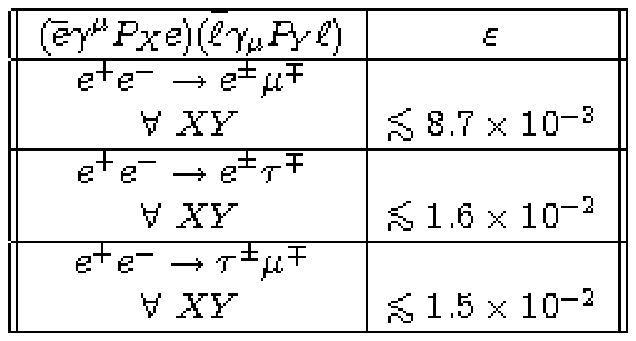}
\vspace{0.5cm}
\includegraphics[width=16pc]{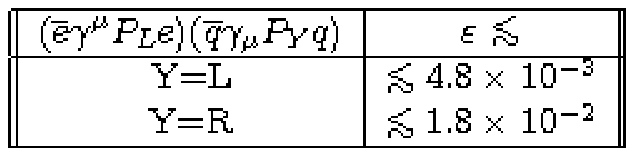}
\end{minipage}
\caption{Estimated bounds on 
$\varepsilon$, from  LEP2, obtained using
the assumptions outlined in the text.
 These could apply to
NSI  models
where the coefficient of dimension six operators
is suppressed by a cancellation. 
\label{tab:LEP2bds}}
\end{figure}

The resulting limits are  listed  in figure
\ref{tab:LEP2bds}.
 Although the bounds
are quoted with two significant figures, they
are merely order of magnitude estimates,
as various constants could appear in the passage
between four-charged-lepton-derivative  operators 
and  NSI operators.

The OPAL experiment saw one $ e^+e^- \to e^\pm \mu^\mp$
 event at $\sqrt{s} = 189-209$ GeV, and published limits
\cite{OPALFV}
on $\sigma( e^+e^- \to e^\pm \mu^\mp, e^\pm \tau^\mp,  \tau^\pm \mu^\mp)$
at LEP2 energies. This analysis allows to set the  bounds
on  $\varepsilon$ for flavour-changing NSI given in 
figure \ref{tab:LEP2bds}.

\section{ LHC discovery reach}
\label{LHC}

As mentioned before, the bound on $\Lambda_8 \lsim 2 TeV$ suggests that the New Physics particles responsible of NSIs would show up as resonances at the LHC, provided the interactions are perturbative. We want to take a more pessimistic approach, and  study the prospects
at the LHC for  NSI involving quarks, in the improbable,
but compartively model-independent,
scenario that NSIs appear as contact
interactions.

In a gauge invariant dimension eight NSI
operator,
the  $H_0^* H_0 \bar{\nu}_\a \nu_\b$
interaction is accompagnied by
$H^+ H^- \bar{e}_\a e_\b$, which
could be expected to reincarnate, after electroweak symmetry
breaking, as
a vertex involving  $W^+ W^- \bar{e}_\a e_\b$.   
This expectation
can be formalised, at energies $\gg m_W$,
via the Equivalence Theorem\cite{EqT}, which 
identifies the Goldstone $H^\pm$
with  the longitudinal  componentof the $W^\pm$.

\begin{figure}[ht]
\begin{minipage}{17pc}
\includegraphics[scale=.35]{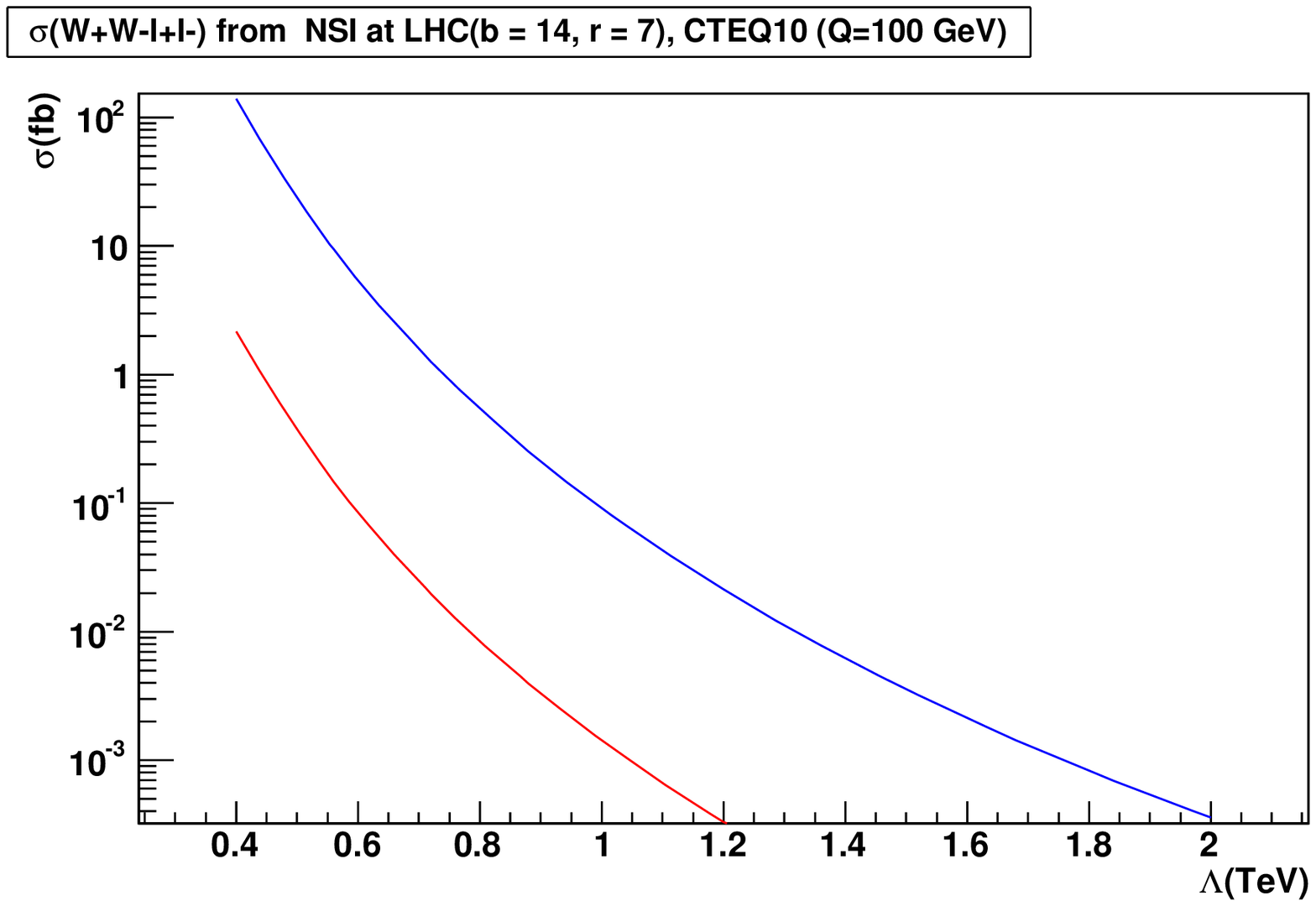}
\caption{ 
$\sigma (pp \to W^+W^- e_\a^+
e^-_\b)$ in fb, at the LHC with 14 TeV or 7 TeV,
due to  a  contact interaction with
coefficient $1/\Lambda_8^4$. 
\label{fig:sigd8} }
\end{minipage}\hspace{3pc}%
\begin{minipage}{17pc}
\includegraphics[scale=.9]{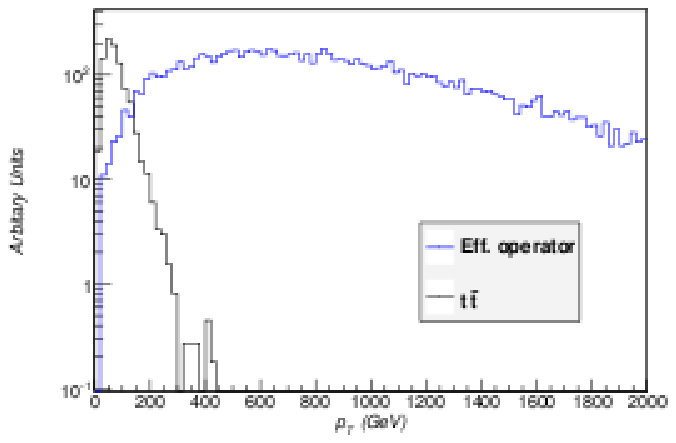}
\caption{ $p_T$ of  leptons at the 14 TeV LHC
from  NSI (flat) and $t \bar{t}$ (peaked),
normalised to fit both lines in the same scale.
\label{fig:pT}}
\end{minipage}
\end{figure}

In the Equivalence Theorem limit,
the partonic cross-section  for $ \bar{q} q \to 
e_\a^+ e_\b^- W^+ W^-$ is 
calculable, and small due to  four body  phase space.
We obtain
$\sigma(p  p \to
H^+H^-e_\a^+e^-_\b)$   plotted in
figure \ref{fig:sigd8}.

To determine the reach of the LHC  in $\epsilon_{\a \b}$,
we need to consider  backgrounds, such as
the $t\bar{t}$ + jets,  and signal selection. 
We
simulated NSI events
 using FeynCalc~\cite{FeynCalc} and MadGraph v5~\cite{MadGraph5}, 
to obtain the $p_T$ distribution
of the signal leptons
 Fig.~\ref{fig:pT}.  We also plot the $p_T$ 
of objects in a $t \bar{t}$  sample. 
The signal and background distributions are well separated, 
so asking for a  $p_T$ cut  of 
order 400 GeV would reduce the $t \bar{t}$ backgrounds
to less 
than $10^{-5}$ fb and keeps 70\% of our signal.
In particular, this suggests
sensitivity to NSI interactions producing taus. 
However, if such high-$p_T$ events were seen, 
boosted taus would be hard to tag, 
although they  may show up as fat jets.

Summarizing, the cross-section is small,
 but the  signal is characterized by well-separated, 
highly boosted objects in a pretty spherical event. 
Using these characteristics, especially a cut on $p_T$, 
we showed that the largest background,
from $t \bar{t}$ can be reduced  below the 
signal. The LHC reach depends on the value of $\varepsilon$, 
but if we assume the NSI signal  is
background-free, and  ask for 100 events at a luminosity ${\cal L}$ 
(in fb$^{-1}$),  then the reach in $\varepsilon$ is
\bea
\varepsilon \sim 3 \times  10^{-2}/\sqrt{{\cal L}}
\label{LHCbds}
\eea
For example, for a luminosity of 100 fb$^{-1}$,  
the  14 TeV LHC  could be sensitive to 
NSI-induced contact interactions corresponding to 
\bea
\varepsilon \gsim  3 \times  10^{-3} ~~~.
\eea


\section*{References}
\begin{thebibliography}{9}





\bibitem{nuFact} see {\it e.g.}  A.~Bandyopadhyay {\it et al.} [ ISS Physics Working Group Collaboration ],
  ``Physics at a future Neutrino Factory and super-beam facility,''
  Rept.\ Prog.\ Phys.\  {\bf 72 } (2009)  106201.
  [arXiv:0710.4947 [hep-ph]]


\bibitem{slepton}
  S.~Antusch, J.~P.~Baumann, E.~Fernandez-Martinez,
  Nucl.\ Phys.\  {\bf B810 } (2009)  369-388.
  [arXiv:0807.1003 [hep-ph]].


\bibitem{Biggio:2009nt}
  C.~Biggio, M.~Blennow, E.~Fernandez-Martinez,
  JHEP {\bf 0908 } (2009)  090
  [arXiv:0907.0097 [hep-ph]]. 


\bibitem{SD}
  S.~Davidson, V.~Sanz,
  ``Non-Standard Neutrino Interactions at Colliders,''
  [arXiv:1108.5320 [hep-ph]].


\bibitem{tdrs}
See for example the ATLAS and CMS Technical Design Reports and references therein.

G.~Aad {\it et al.}  [The ATLAS Collaboration],
  arXiv:0901.0512 [hep-ex].
 G.~L.~Bayatian {\it et al.}  [CMS Collaboration],
  J.\ Phys.\ G {\bf 34} (2007) 995.


\bibitem{Gavela}
  M.~B.~Gavela, D.~Hernandez, T.~Ota, W.~Winter,
  Phys.\ Rev.\  {\bf D79 } (2009)  013007.
  [arXiv:0809.3451 [hep-ph]].


\bibitem{bourilkov}
 D.~Bourilkov,
  Phys.\ Rev.\  D {\bf 64} (2001) 071701
  [arXiv:hep-ph/0104165].


\bibitem{ALEPH}
  S.~Schael {\it et al.}  [ALEPH Collaboration],
  Eur.\ Phys.\ J.\  C {\bf 49} (2007) 411
  [arXiv:hep-ex/0609051].


\bibitem{OPAL}
  G.~Abbiendi {\it et al.}  [OPAL Collaboration],
  Eur.\ Phys.\ J.\  C {\bf 33} (2004) 173
  [arXiv:hep-ex/0309053].




\bibitem{OPALFV}
  G.~Abbiendi {\it et al.} [ OPAL Collaboration ],
  Phys.\ Lett.\  {\bf B519 } (2001)  23-32.
  [hep-ex/0109011].


\bibitem{EqT}
  J.~M.~Cornwall, D.~N.~Levin, G.~Tiktopoulos,
  Phys.\ Rev.\  {\bf D10 } (1974)  1145.
 B.~W.~Lee, C.~Quigg, H.~B.~Thacker,
  Phys.\ Rev.\  {\bf D16 } (1977)  1519.
 M.~S.~Chanowitz, M.~K.~Gaillard,
  Nucl.\ Phys.\  {\bf B261 } (1985)  379.
   H.~G.~J.~Veltman,
  Phys.\ Rev.\  {\bf D41 } (1990)  2294.



\bibitem{FeynCalc}
See http://www.feyncalc.org/ for documentation.


\bibitem{MadGraph5}
 J.~Alwall, M.~Herquet, F.~Maltoni, O.~Mattelaer, T.~Stelzer,
  JHEP {\bf 1106}, 128 (2011).
  [arXiv:1106.0522 [hep-ph]].




\end{thebibliography}
\end{document}